\documentclass[preprint,12pt]{elsarticle}

\usepackage[usenames,dvipsnames]{xcolor}
\usepackage{amssymb}

\begin{document}

\begin{frontmatter}



\title{Photoalignment at the nematic liquid crystal - polymer interface: the importance of the liquid crystalline molecular structure}


\author{Ameer R.K. Nassrah}

\author{Istv\'an J\'anossy}

\author{Tibor T\'oth-Katona\corref{1}}
\cortext[1]{Corresponding author.}
\ead{tothkatona.tibor@wigner.hu}

\address{Institute for Solid State Physics and Optics, Wigner Research Centre for Physics, P.O. Box 49, H-1525 Budapest, Hungary.}

\begin{abstract}
The influence of the molecular structure of the nematic liquid crystal (NLC) on the photoalignment process at the NLC -- polymer interface has been investigated experimentally.
NLCs having exclusively phenyl-, or cyclohexane rings in the rigid core, as well as NLCs containing both phenyl and cyclohexane groups have been considered.
Substantial differences have been found in the photoalignment process depending on the molecular structure of the NLC, supporting the assumption that the polymer -- NLC
interface should be regarded as a coupled system, where the two components mutually influence each other. A phenomenological explanation is given for the observed differences.

\end{abstract}


\end{frontmatter}


\section{Introduction}
\label{Intro}

Photoalignment of nematic liquid crystals (NLCs) has been discovered almost three decades ago \cite{Gibbons1991,Dyadyusha1992,Gibbons1995}. The process is not only an alternative to the standard aligning methods (e.g., mechanical rubbing of polyimide layers) ensuring the correct operation of devices based on NLCs, but also enables the reorientation of the liquid crystal director {\bf n} in a contactless manner through light irradiation \cite{Ube2017}.

Among others, photoalignment can be achieved by the trans-cis (E/Z) isomerization of azo dyes \cite{Rau1990} coated on the substrate either as a molecular monolayer \cite{Aoki1992,Yi2009,Janossy2011,Janossy2014,Janossy2014a}, or as a polymer film
\cite{Gibbons1991,Gibbons1995,Ichimura1989,Ichimura1993,Akiyama1995,Janossy1999,Janossy2001,Palffy2002}.
Photo-reorientation process on samples with a photosensitive molecular monolayer has been compared recently to that on samples with a photosensitive polymer layer, and considerable differences have been found \cite{Janossy2018}.
Typically, the liquid-crystal (LC) sandwich cell for photoalignment is constructed with one photosensitive substrate and one traditionally prepared reference plate (e.g., with rubbed polyimide layer). In some cases, the light irradiation creates cis (Z) isomers, triggering a homeotropic-to-planar transition ‒ often called 'out-of-plane alignment' photocontrol \cite{Aoki1992,Ichimura1993,Kawanishi1991}. In the other, more often used mechanism, the liquid crystal molecules remain in the plane of the substrate ('in-plane alignment' photocontrol), and the azimuthal angle of the director is controlled with polarized light: after the irradiation the director becomes perpendicular to the light polarization direction \cite{Gibbons1991,Gibbons1995,Ichimura1993,Akiyama1995,Palffy2002}. The photocontrol mechanisms, as well as the most common photosensitive materials are well described in a review on photoalignment of NLCs \cite{Ichimura2000}.

We have recently provided experimental evidence that the standard two-dimensional description of photoalignment is insufficient in a number of cases \cite{Tothkatona2019}. Besides the azimuthal photo-reorientation, zenithal photoalignment, as well as temperature induced anchoring transition have also been observed. Moreover, these measurements have shown that the polymer-liquid crystal interface should be regarded as a coupled system, where the two components mutually influence each other, as it was suggested earlier \cite{Janossy2001}.
The coupling between the polymer and the liquid crystal has also been considered via the so called "gliding of the easy axis" \cite{Vilfan2001,Vilfan2003,Joly2004,Pasechnik2006,Oswald2008}, or via the correlation between the surface and the bulk orientational properties of LCs in various geometries \cite{Emelyanenko2011,Kalinin2017}.

In this work we focus on the influence of the liquid crystalline molecular structure on the photo-reorientation process, a task which was on the top of the list to do in the future, envisioned in Ref. \cite{Tothkatona2019}. We have compared  photoalignment of nematic liquid crystals with different rigid cores on the same photosensitive layer. Namely, we have repeated the previous measurements on NLCs with two phenyl rings; additionally, we have investigated materials with one phenyl and one cyclohexane ring, as well as that having two cyclohexane rings. Notable differences have been found in the photo-reorientation process in these materials, for which we give a phenomenological explanation based on the presence or absence of the $\pi - \pi$ aromatic interactions between the NLC and the polymer.

\section{Experimental details}

\label{Exp}

\subsection{Materials and sample preparation}

\begin{figure}
\begin{center}
\includegraphics[width=20pc]{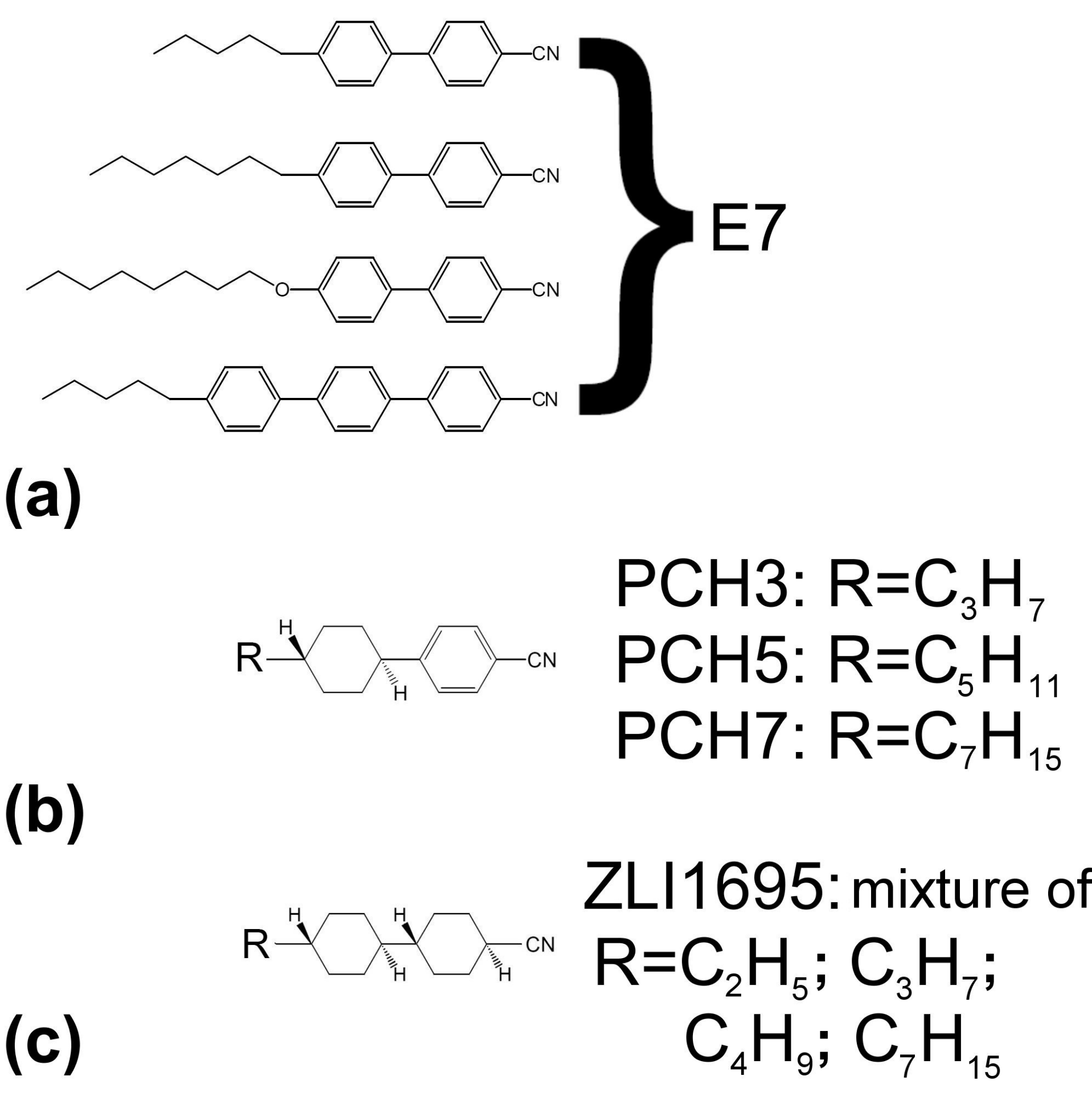}
\caption{Molecular structure of NLCs: (a) only phenyl rings in the rigid core (mixture E7); (b) phenyl and cyclohexane groups in the rigid core (PCH3, PCH5, PCH7); (c) only cyclohexane rings in the rigid core (mixture ZLI1695).} \label{Nassrah_Fig1}
\end{center}
\end{figure}

The systematic variation of the rigid core of NLCs has been ensured by the choice of mesogenic materials. For the NLC having only phenyl rings in the rigid core, the representative NLC mixture E7 has been selected [Fig. \ref{Nassrah_Fig1}(a)], for which most of the physical parameters (and their temperature dependence) are known. Though we have already reported the results on the azimuthal and zenithal photo-reorientation, and on the temperature induced anchoring transition for this NLC \cite{Tothkatona2019}, some of the photoaligning properties have been re-measured and will be discussed in this work. The nematic-to-isotropic phase transition temperature $T_{NI}$ (clearing point) for E7 has been found at around $60^{\circ}$C.
For NLCs containing both phenyl and cyclohexane rings some members of the PCH homologous series have been chosen [Fig. \ref{Nassrah_Fig1}(b)]. The clearing point of these compounds have been found at $47.0^{\circ}$C, $53.5^{\circ}$C and $58.5^{\circ}$C for PCH3, PCH5 and PCH7, respectively. For the NLC having exclusively cyclohexane rings in the rigid core, the mixture ZLI1695 [Fig. \ref{Nassrah_Fig1}(c)] has been selected, having the nematic phase down to room temperature, with a clearing point of $T_{NI}=72.5^{\circ}$C. One has to note here that all NLC materials have the same polar head, and (in particular for PCH series) somewhat different length of the flexible alkyl-chain.

\begin{figure}
\begin{center}
\includegraphics[width=15pc]{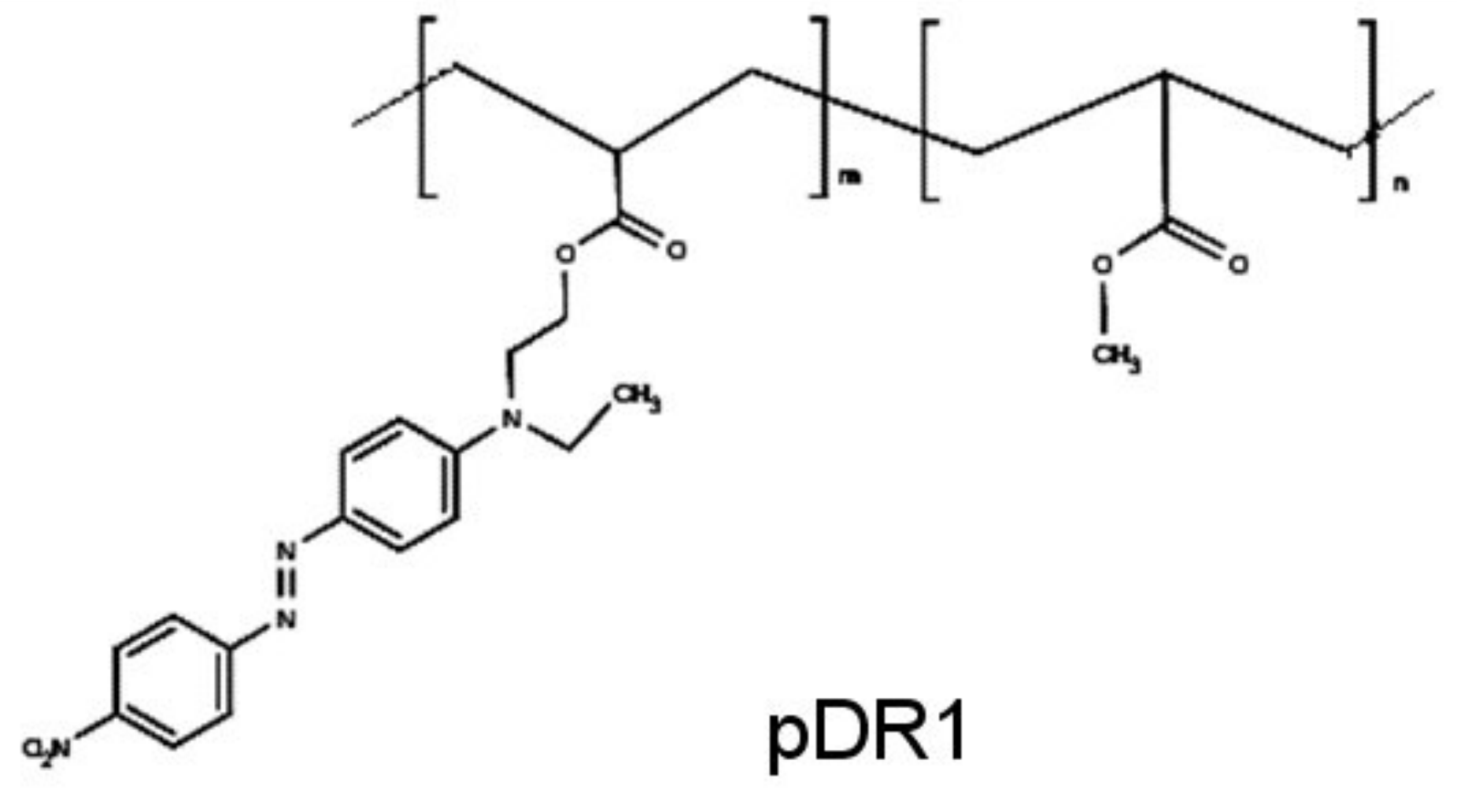}
\caption{Molecular structure of the photosensitive polymer pDR1.} \label{Nassrah_Fig2}
\end{center}
\end{figure}

The photosensitive substrate of the samples has been prepared by spin-coating of polymer polymethyl-methacrilate functionalized with the azo-dye Disperse Red 1 (pDR1 -- see Fig. \ref{Nassrah_Fig2}) as described in \cite{Tothkatona2019} in details. Rubbed polyimide coated slides from E.H.C. Co. (Japan) have been used as the other, reference plates of LC cells. The thickness of the LC cells (assembled with spacers) has been measured by the interferometric method.

Before filling the NLC into the cells, the cell was illuminated with white light from the side of the photosensitive substrate, polarized perpendicularly to the rubbing direction on the reference plate. This procedure ensured a good quality planar initial alignment of the nematic liquid crystal.

\subsection{Experimental setup}

Measurements on the photoalignment and photo-reorientation have been performed on a pump-probe optical setup described in details in Ref. \cite{Janossy2018}.
The polarized pump beam from a DPSS laser ($\approx 20$mW, $\lambda=457$nm) entered the cell from the photosensitive side, defocused to a spot size of few mm, much larger than the diameter of the probe beam. The polarized probe beam from a He-Ne laser ($5$mW, $\lambda=633$nm) was sent through the cell, entering it at the reference plate. Behind the sample the probe beam was sent through a rotating polarizer and its intensity was detected by a photodiode; the signal was connected to a lock-in amplifier. The setup provides the phase and the amplitude of the probe beam transmitted through the sample from which the photo-induced twist angle $\varphi$ can be determined, and information about the zenithal (out-of-plane) photoalignment can be obtained with arrangements described in Ref. \cite{Tothkatona2019}.

\section{Results}
\label{Results}

\subsection{Azimuthal (in-plane) photoalignment}

In the measurements of the in-plane (azimuthal) photoalignment the polarization of both the pump and the probe beam has been set parallel with the initial {\bf n}, and the phase of the probe beam has been measured. The temperature of the samples has been varied from room temperature up to $T_{NI}$. Under these conditions the pump beam is expected to induce a twist deformation at the photosensitive substrate resulting in a twisted LC cell from the initially planar one. In the case of a perfect azimuthal reorientation the twist angle $\varphi$ should saturate at $90^{\circ}$.

\begin{figure}
\begin{center}
{\bf (a)}\includegraphics[width=20pc]{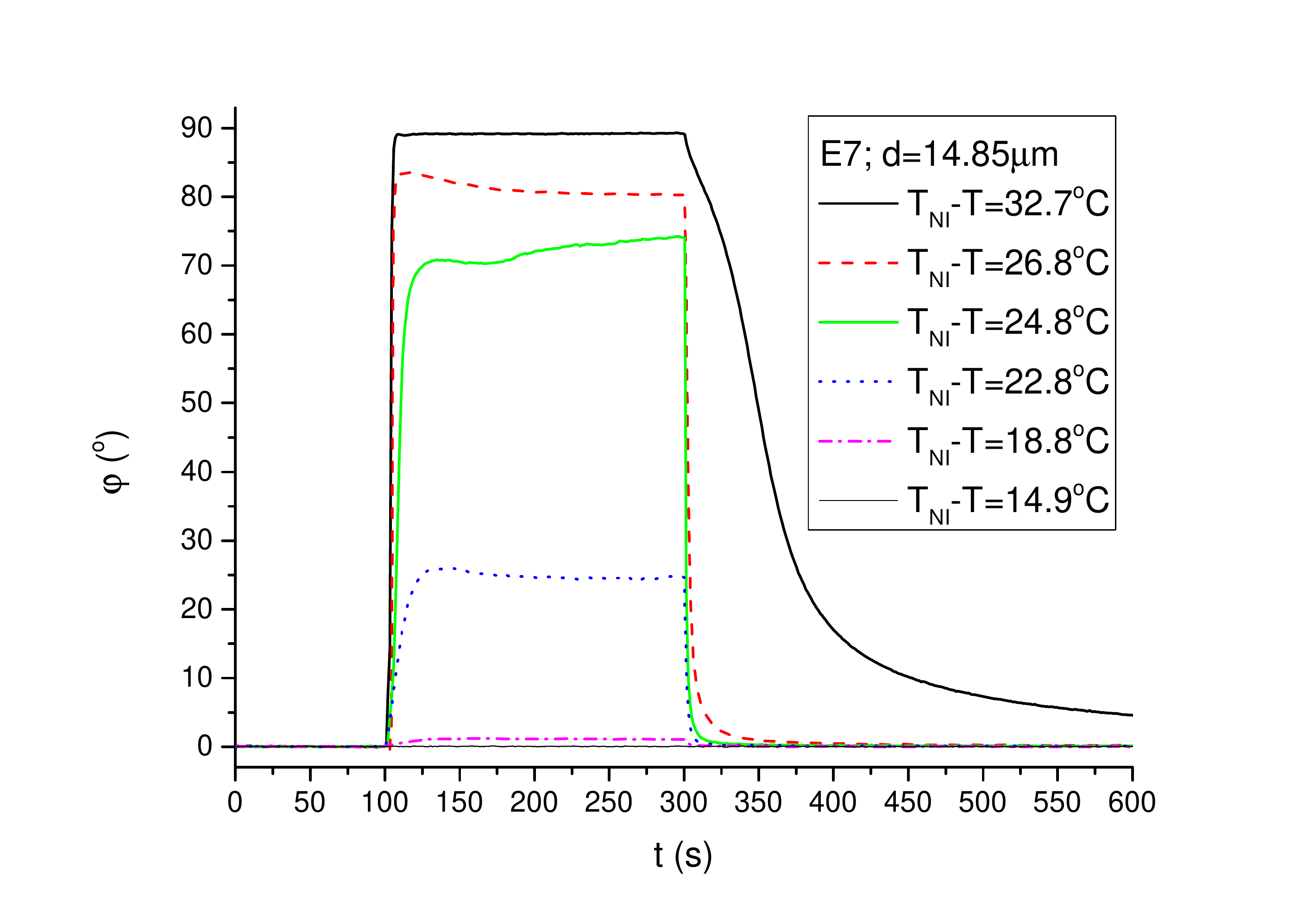} {\bf (b)}\includegraphics[width=20pc]{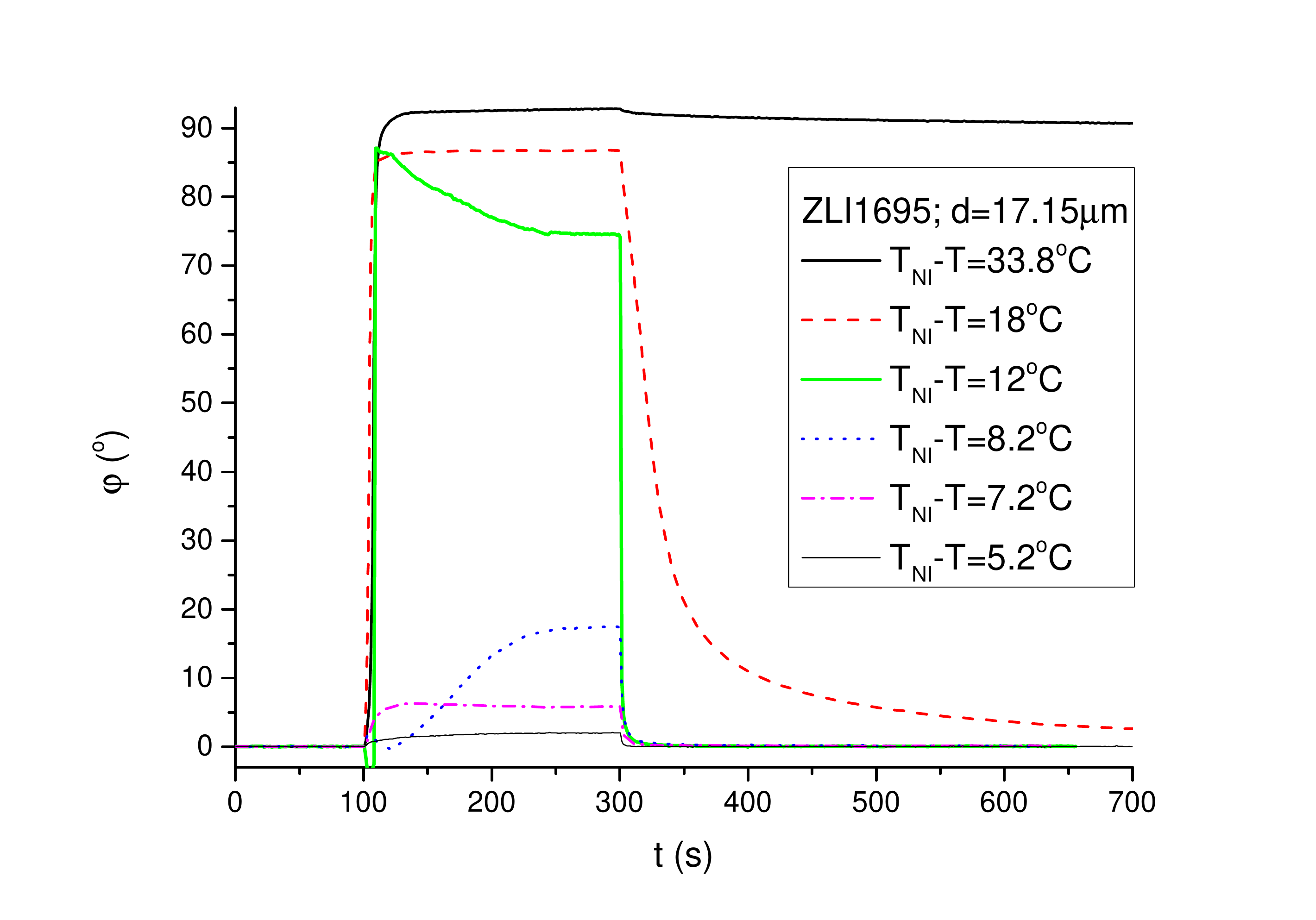}
\caption{Azimuthal photo-reorientation angle $\varphi$ in time, measured at different temperatures in cells filled with NLCs having: (a) phenyl rings in the rigid core (E7), and (b) cyclohexane groups in the rigid core (ZLI1695). The pump-beam was switched on at $t=100$s, and switched off at a time of 300s.} \label{Nassrah_Fig3}
\end{center}
\end{figure}

Figure \ref{Nassrah_Fig3} shows the temporal evolution of the photoinduced twist angle $\varphi$ for LC cells filled with E7 (a) and ZLI1695 (b), measured at different temperatures
$\Delta T = T_{NI}-T$. The pump-beam has been switched on at $t=100$s and switched off at $t=300$s in all measurements. As one sees, at low (room) temperature the azimuthal twist deformation saturates at $\varphi \approx 90^\circ$ [see Fig.~\ref{Nassrah_Fig3}(a) and (b)] for both E7 and ZLI1695, indicating a complete azimuthal photo-reorientation.
With the increase of the temperature, the twist angle gradually decreases, and vanishes for E7 far below $T_{NI}$ [Fig. \ref{Nassrah_Fig3}(a)], while for ZLI1695 just below $T_{NI}$ [Fig. \ref{Nassrah_Fig3}(b)].

\begin{figure}
\begin{center}
{\bf (a)}\includegraphics[width=20pc]{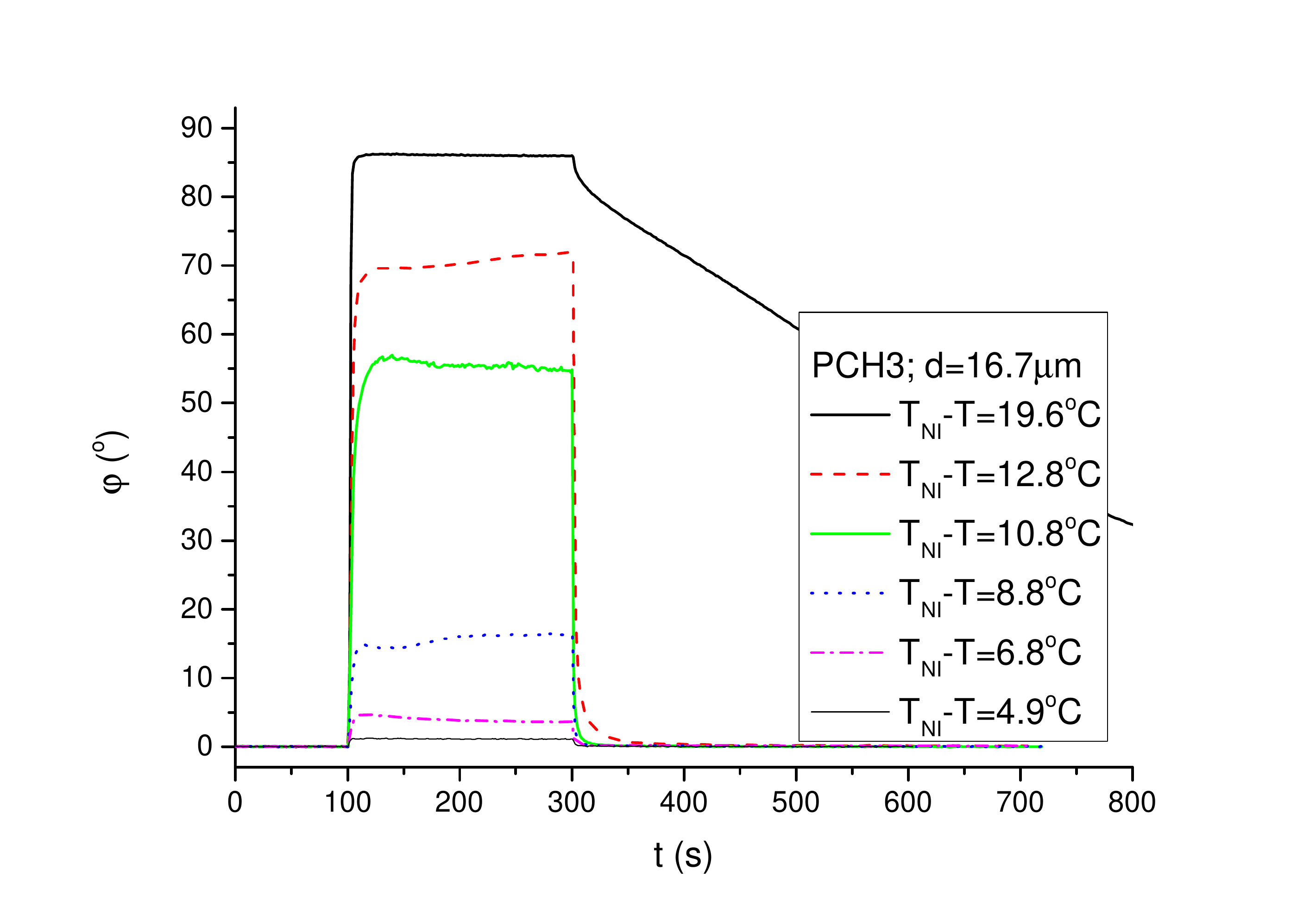} {\bf (b)}\includegraphics[width=20pc]{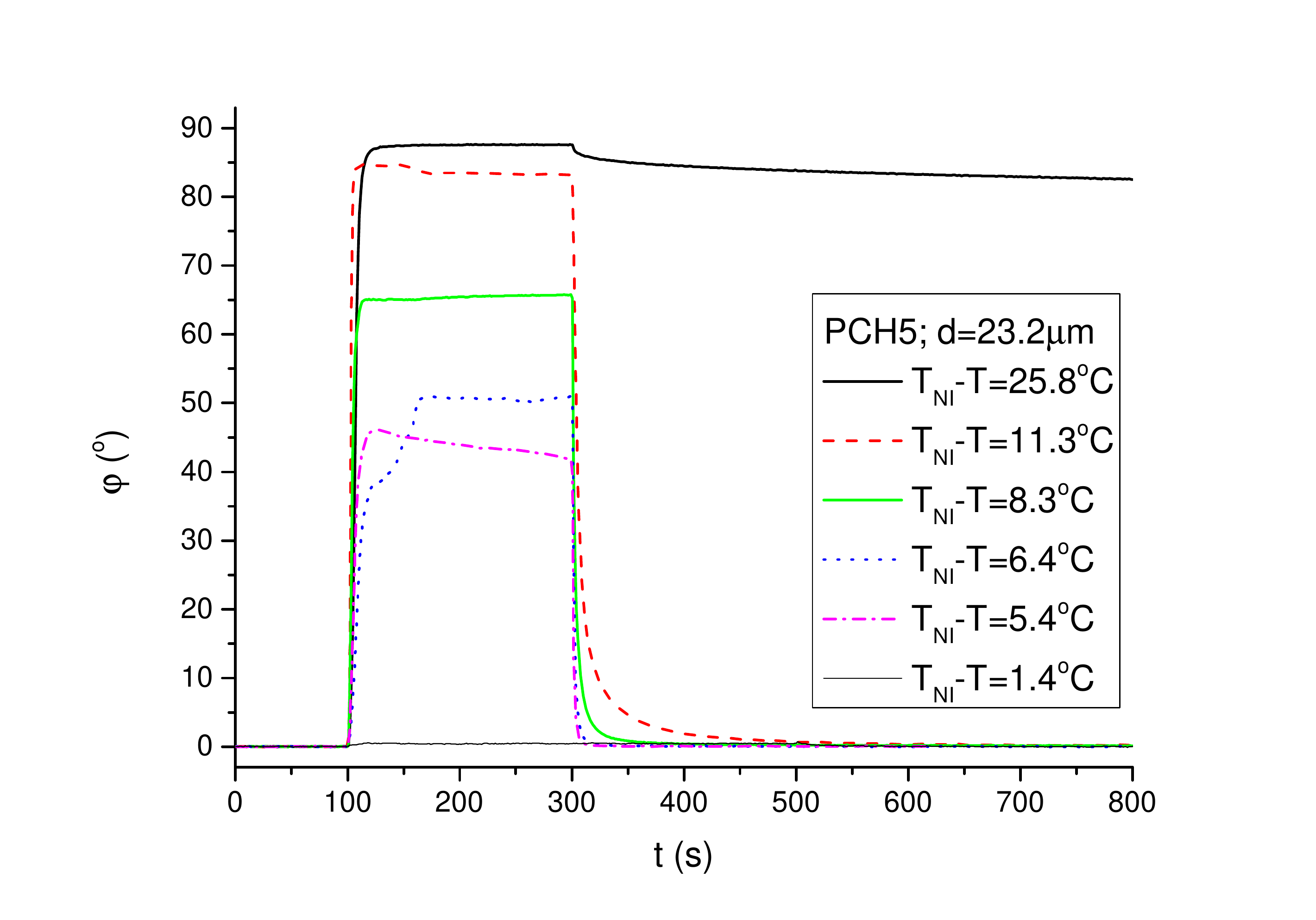} {\bf (c)}\includegraphics[width=20pc]{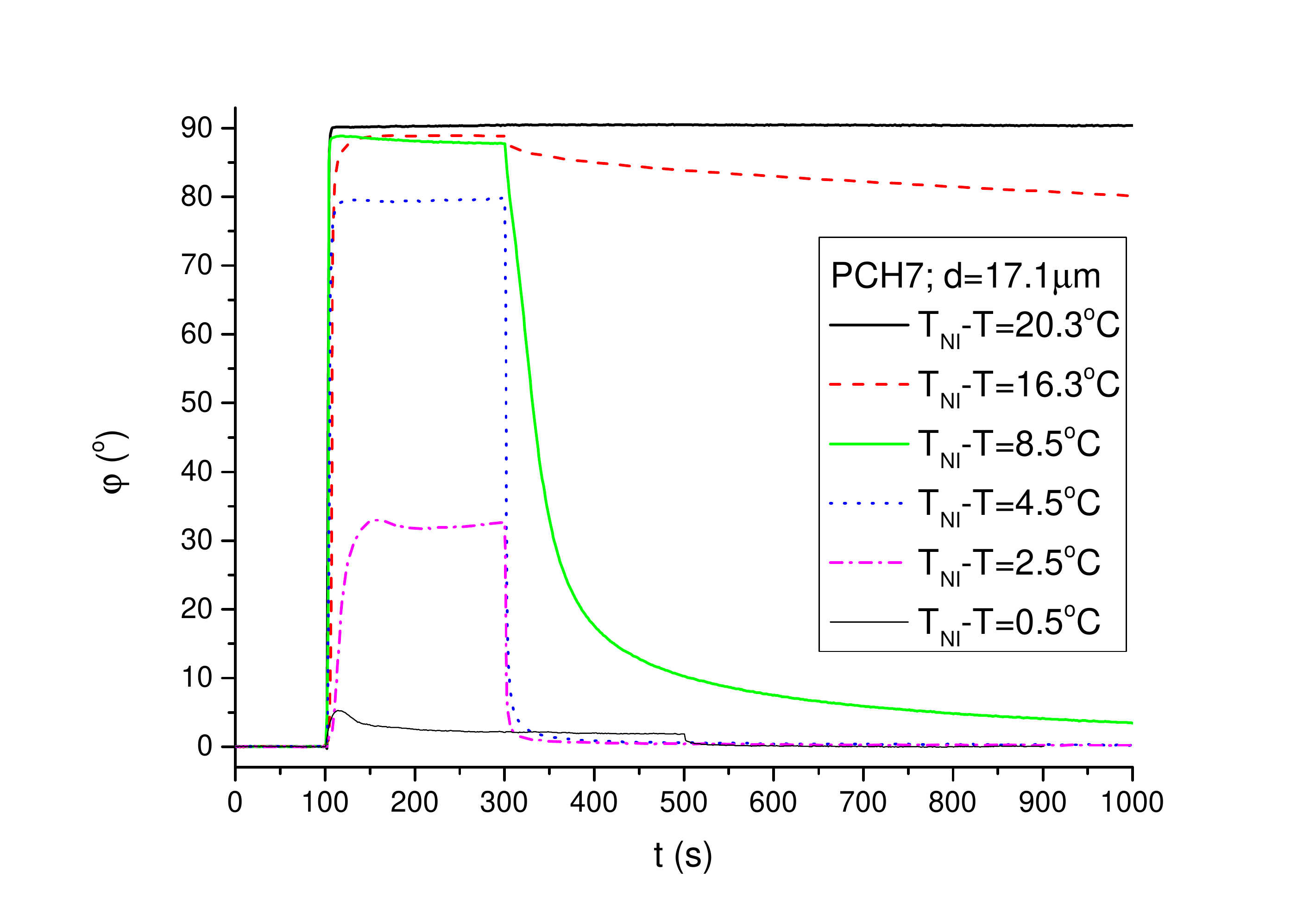}
\caption{Azimuthal photo-reorientation angle $\varphi$ in time, measured at different temperatures in cells filled with NLCs having a phenylcyclohexane rigid core: (a) PCH3, (b) PCH5, and (c) PCH7. The pump-beam was switched on at $t=100$s, and switched off at a time of 300s, or (at the highest temperature for PCH7) at 500s.} \label{Nassrah_Fig4}
\end{center}
\end{figure}

The temporal evolution of $\varphi (t)$ for LC cells filled with NLCs from PCH series is shown in Fig. \ref{Nassrah_Fig4}. The results on the photoinduced twist angle practically correspond to those obtained on the cell filled with ZLI1695.

\begin{figure}
\begin{center}
\includegraphics[width=20pc]{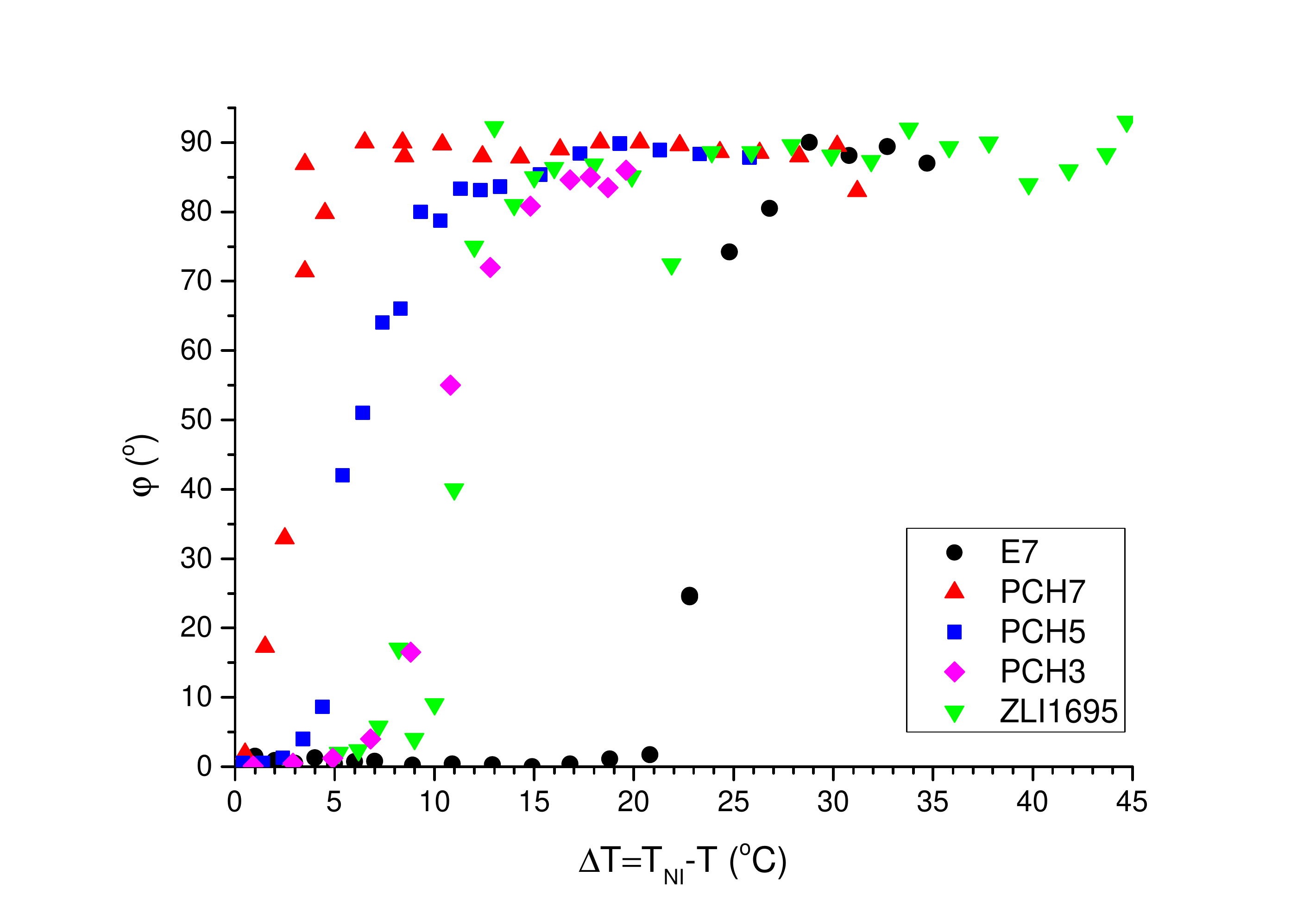}
\caption{Temperature dependence of the azimuthal photo-reorientation angle $\varphi$ measured in cells filled with various NLCs (E7, PCH5, PCH7 and ZLI1695) as indicated in the legend.} \label{Nassrah_Fig5}
\end{center}
\end{figure}

In Fig. \ref{Nassrah_Fig5} we present the temperature dependence of the saturated twist angle measured for E7, ZLI1695 and PCH series. For all NLCs a sudden decrease of $\varphi$ has been observed  below $T_{NI}$. However, while the decrease for E7 occurs in the temperature range of $T_{NI}-T = 20-25^{\circ}$C (in full agreement with the results published earlier
\cite{Tothkatona2019}), for ZLI1695 and for PCH NLCs it is much closer to the nematic-to-isotropic phase transition temperature ($T_{NI}-T \lesssim 10^{\circ}$C). Moreover, concentrating on PCH series in Fig. \ref{Nassrah_Fig5}, the decrease of $\varphi$ occurs closer to $T_{NI}$ as the length of the alkyl chain is increased, which may indicate a possible influence of the flexible alkyl chain on the photoalignment.

\subsection{The magnitude of the zenithal (out-of-plane) photoalignment}

In samples with E7, having biphenyl in the rigid core, a considerable zenithal photoalignment has been measured previously \cite{Tothkatona2019}, in the temperature range above which the azimuthal photo-reorientation angle $\varphi$ suddenly drops (i.e., for $\Delta T = T_{NI}-T \lesssim 20^{\circ}$C -- see also in Fig. \ref{Nassrah_Fig5}).
To test whether a similar zenithal photoalignment occurs in samples with PCH compounds, or with ZLI1695, the experimental setup for zenithal reorientation was used:
the polarization of the probe beam enclosed $45^{\circ}$ with {\bf n}, and the amplitude of the signal was detected.
Measurements have been performed with the polarization of the pump beam perpendicular to the initial {\bf n}, in which geometry no azimuthal photoalignment is expected.
With this setup, if a significant out-of-plane photoalignment occurs, oscillations in the transmitted light intensity of the probe beam should appear, as in the measurements on the electric-, or magnetic-field induced Fr\'eedericksz transition -- see e.g., Ref. \cite{Parshin2014}.

\begin{figure}
\begin{center}
\includegraphics[width=20pc]{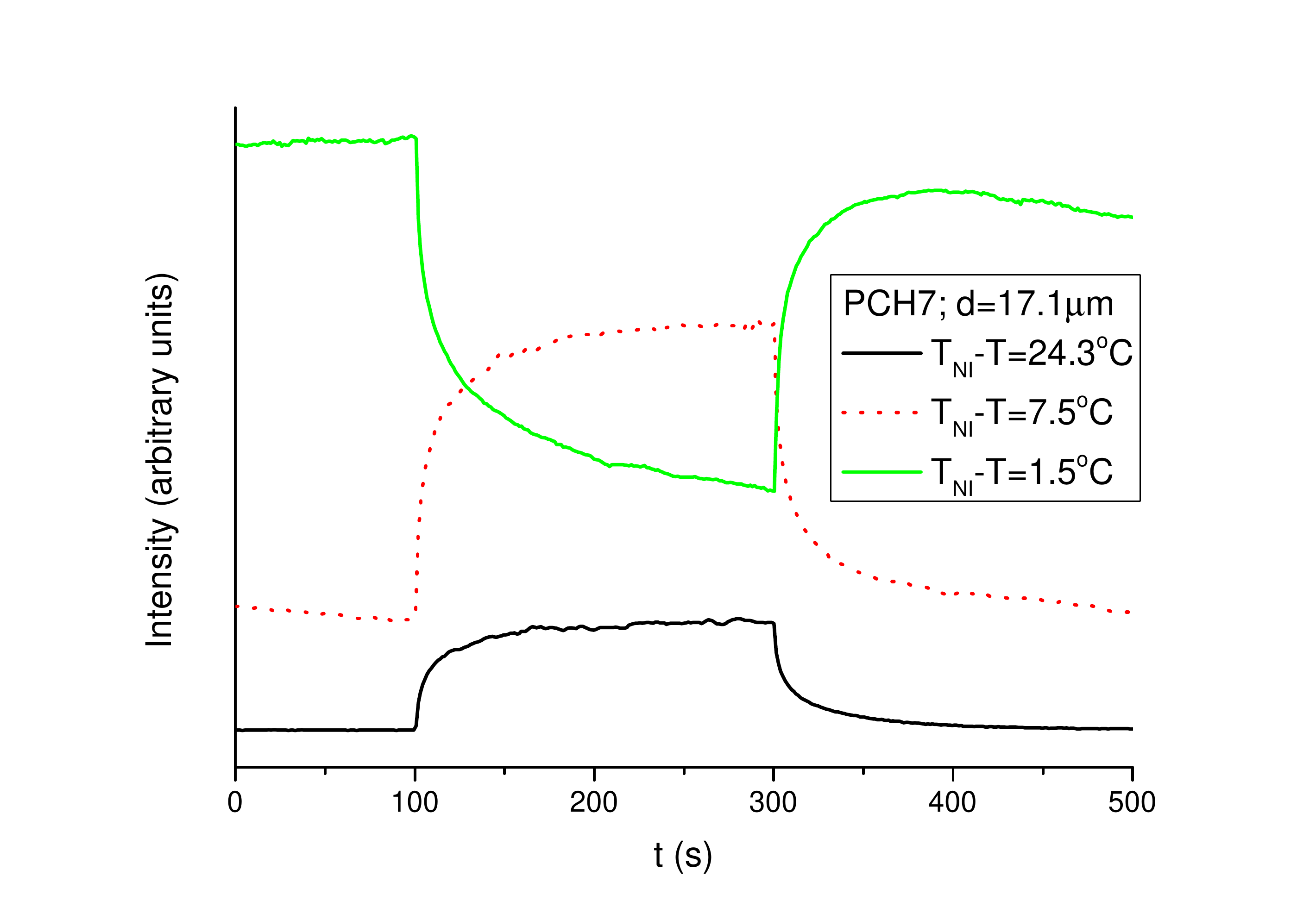}
\caption{Temporal variation of the transmitted light intensity of the probe beam measured in a PCH7 cell at different temperatures in the setup for detection of zenithal
photo-reorientation (pump beam polarization perpendicular to {\bf n}, probe beam polarization encloses $45^{\circ}$ with {\bf n}).} \label{Nassrah_Fig6}
\end{center}
\end{figure}

Measurements on the zenithal photoalignment have been performed from room temperature to the clearing temperature $T_{NI}$ for PCH3, PCH5, PCH7, as well as for ZLI1695.
None of these measurements resulted in oscillations of the transmitted light intensity, indicating the absence of a significant zenithal photo-reorientation, in contrast to the samples filled with E7, in which such oscillations were clearly detected -- see in Ref. \cite{Tothkatona2019}. Instead of oscillations, in all PCH and ZLI1695 samples, a slight change in the transmitted light intensity has only been observed at all temperatures, which may originate either from a small misalignment of the director at the two bounding surfaces, or from a small misalignment of the polarization direction of the pump beam and {\bf n}, or from a slight zenithal photoalignment as pointed out in Ref. \cite{Tothkatona2019}.
As a representative example of the small changes in the light intensity, in Fig. \ref{Nassrah_Fig6} we give the temporal variation of the light intensity in a $d=17.1\mu$m thick cell filled with PCH7 for three temperatures $\Delta T=T_{NI}-T=24.3^{\circ}$C, $7.5^{\circ}$C, and $1.5^{\circ}$C. The pump beam has been switched on at $t=100$s, and switched off at $t=300$s. Based on these measurements one can conclude that in samples with PCH NLCs and with ZLI1695 mixture, the magnitude of the zenithal photoalignment is either zero, or it is much smaller than that detected in the NLC with biphenyl rigid core (E7) \cite{Tothkatona2019}.

\subsection{Director back-relaxation from the azimuthal photo-reorientation}

In our previous work \cite{Tothkatona2019}, we have briefly discussed the director back-relaxation process in NLCs  with biphenyl rigid core (E7, E63 and 5CB). The temperature dependence of the back-relaxation dynamics has been explained by the temperature dependence of the equilibrium concentration of the trans-cis isomers, and by that of the cis-to-trans isomerization kinetics. It has been also noted, that besides the temperature dependence of the isomerization kinetics, the back-relaxation speed also depends on the NLC material that is put in contact with the polymer [{\it cf.} Figs. 1(a) and 1(b) of Ref. \cite{Tothkatona2019} for NLCs E7 and E63, respectively]. Though the back-relaxation in E63 has been found slower than that in E7, at a given relative temperature $T_{NI}-T$ the relaxation times were of the same order of magnitude.

\begin{figure}
\begin{center}
\includegraphics[width=20pc]{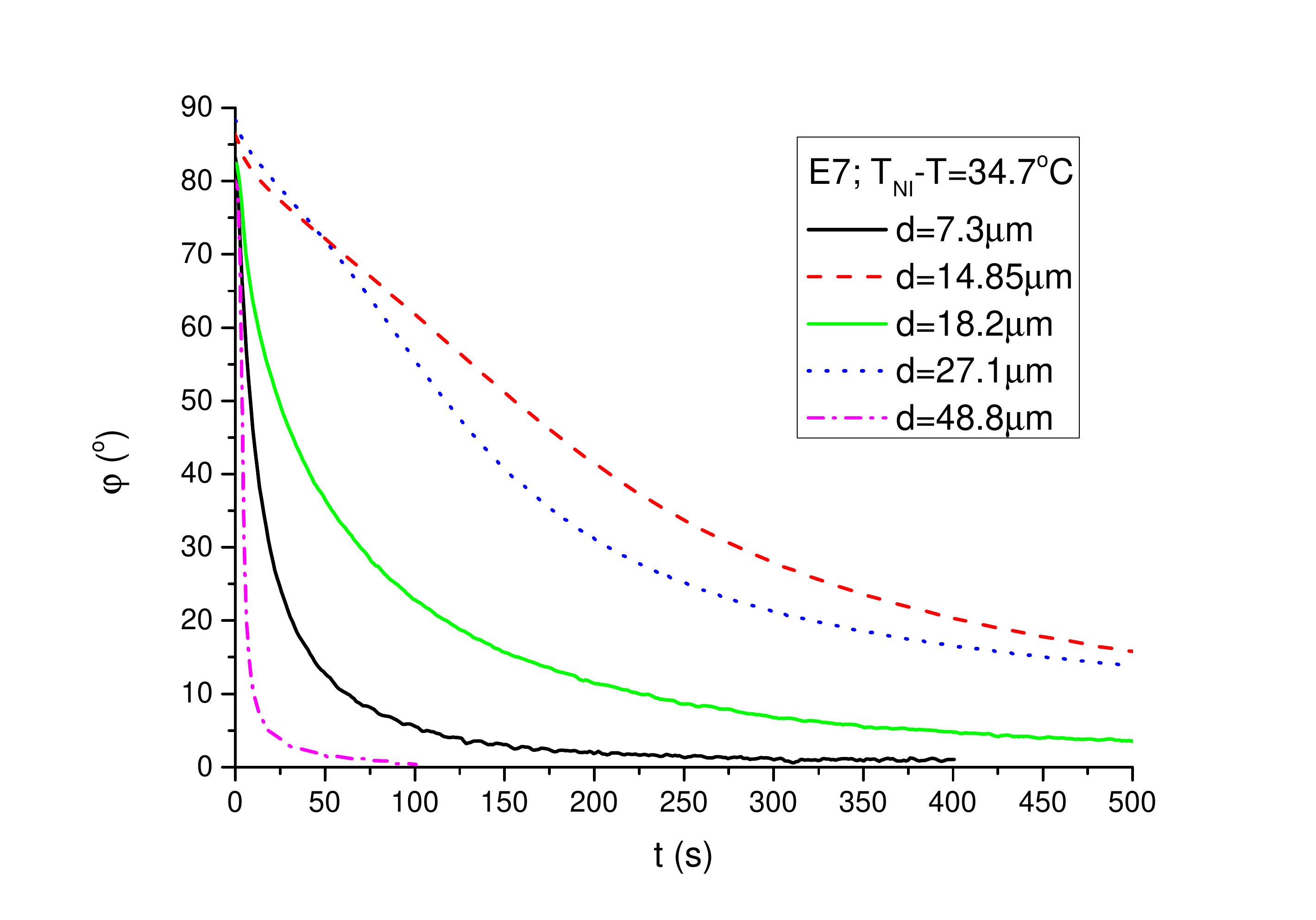}
\caption{Back-relaxation of the azimuthal photo-reorientation angle $\varphi$ (pump beam switched off at $t=0$s) measured at a given temperature in cells of different thickness $d$ filled with E7 as indicated in the legend.} \label{Nassrah_Fig7}
\end{center}
\end{figure}

Here, first we examine the thickness dependence of the back-relaxation process. Fig. \ref{Nassrah_Fig7} shows the temporal evolution of the back-relaxation of the photoinduced azimuthal (twist) angle $\varphi$ in cells of different thickness $d$, filled with E7 and measured at the same temperature $T_{NI}-T = 34.7^{\circ}$C. Surprisingly, no straightforward thickness dependence of the relaxation kinetics has been found: the thickest samples relaxed the fastest, followed by thinnest cell. This finding is reminiscent of non-trivial $d$ dependence of the temperature range at which the sudden decrease of $\varphi$ is detected for E7 \cite{Tothkatona2019}.

\begin{figure}
\begin{center}
{\bf (a)}\includegraphics[width=20pc]{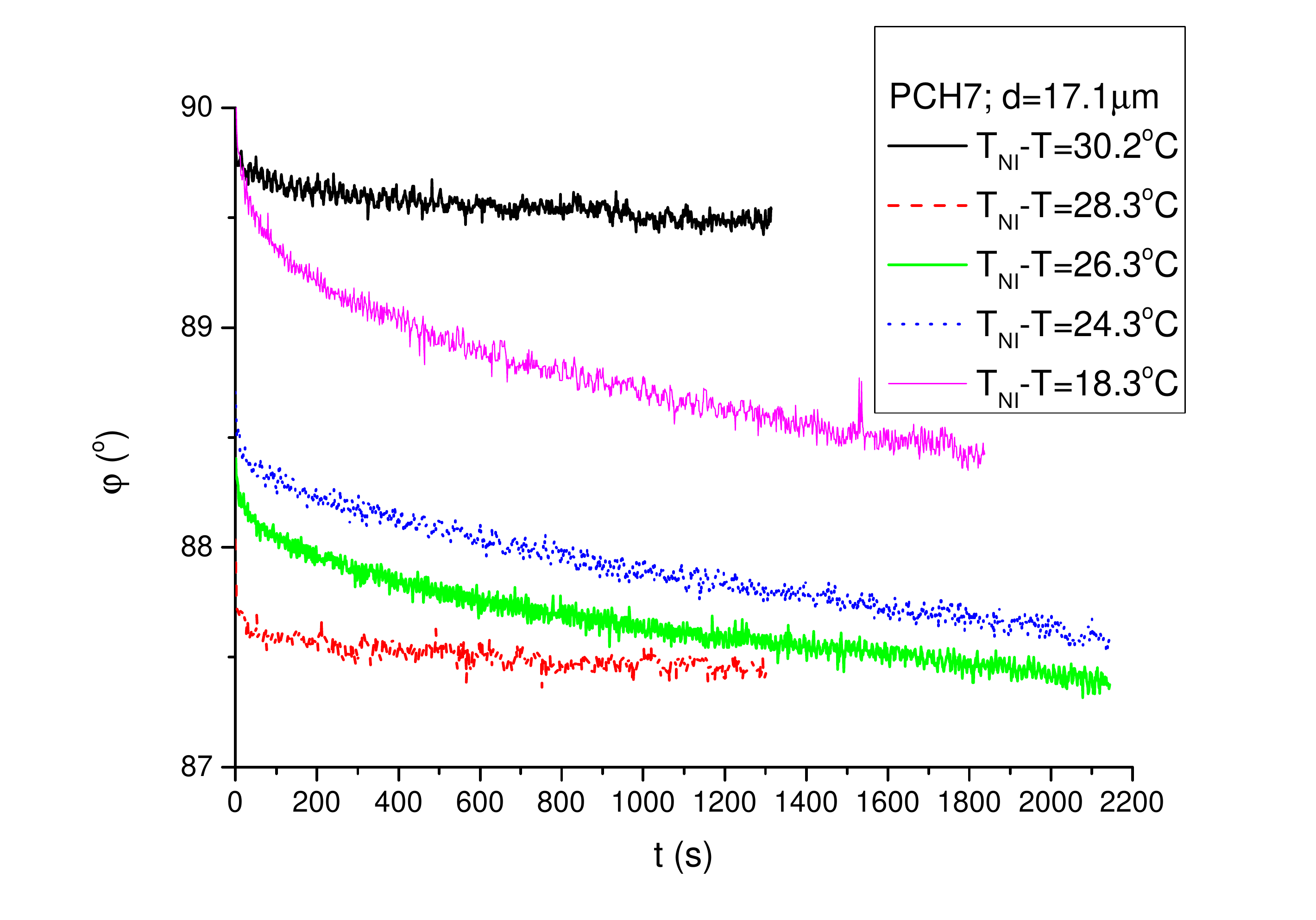} {\bf (b)}\includegraphics[width=20pc]{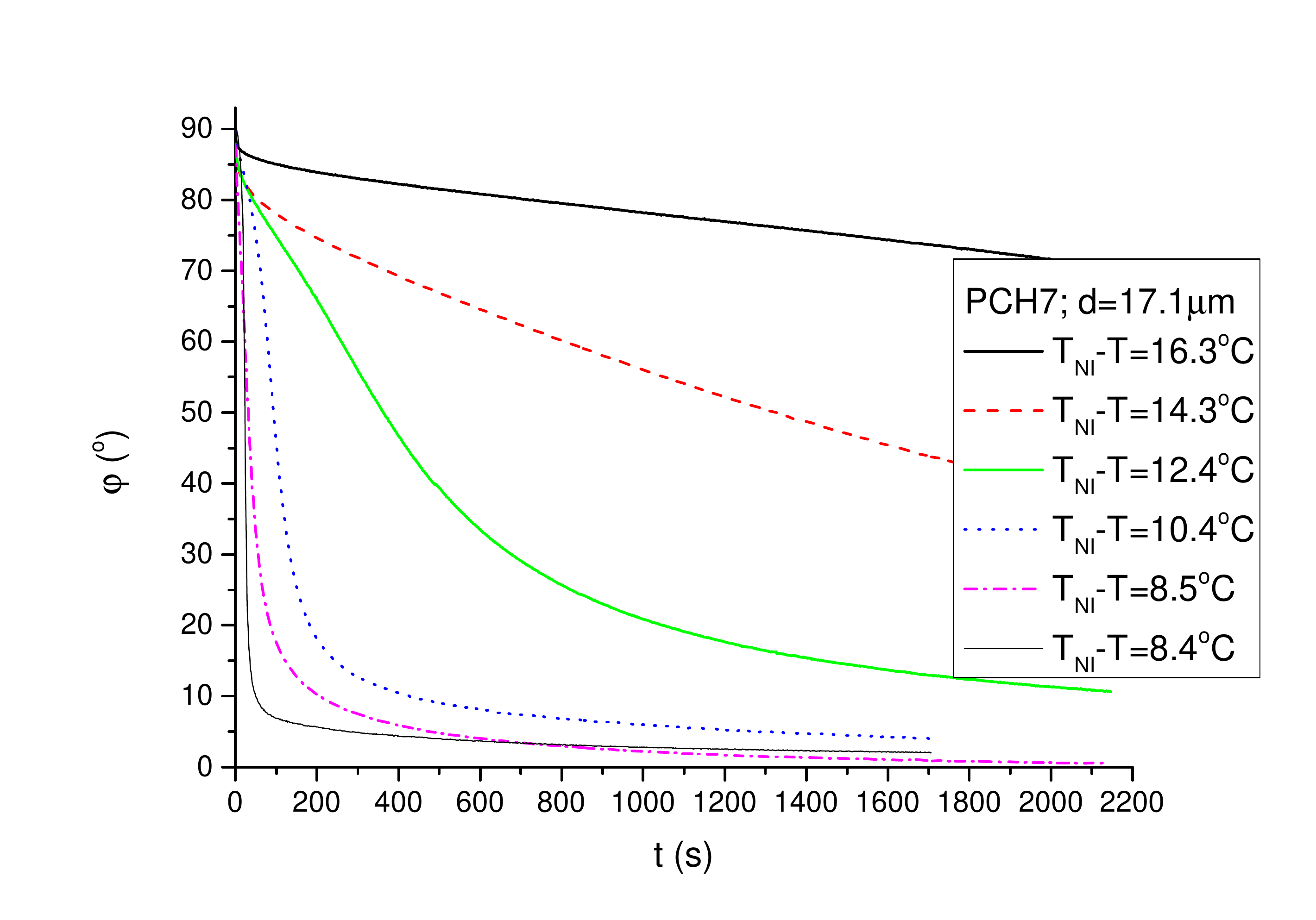} {\bf (c)}\includegraphics[width=20pc]{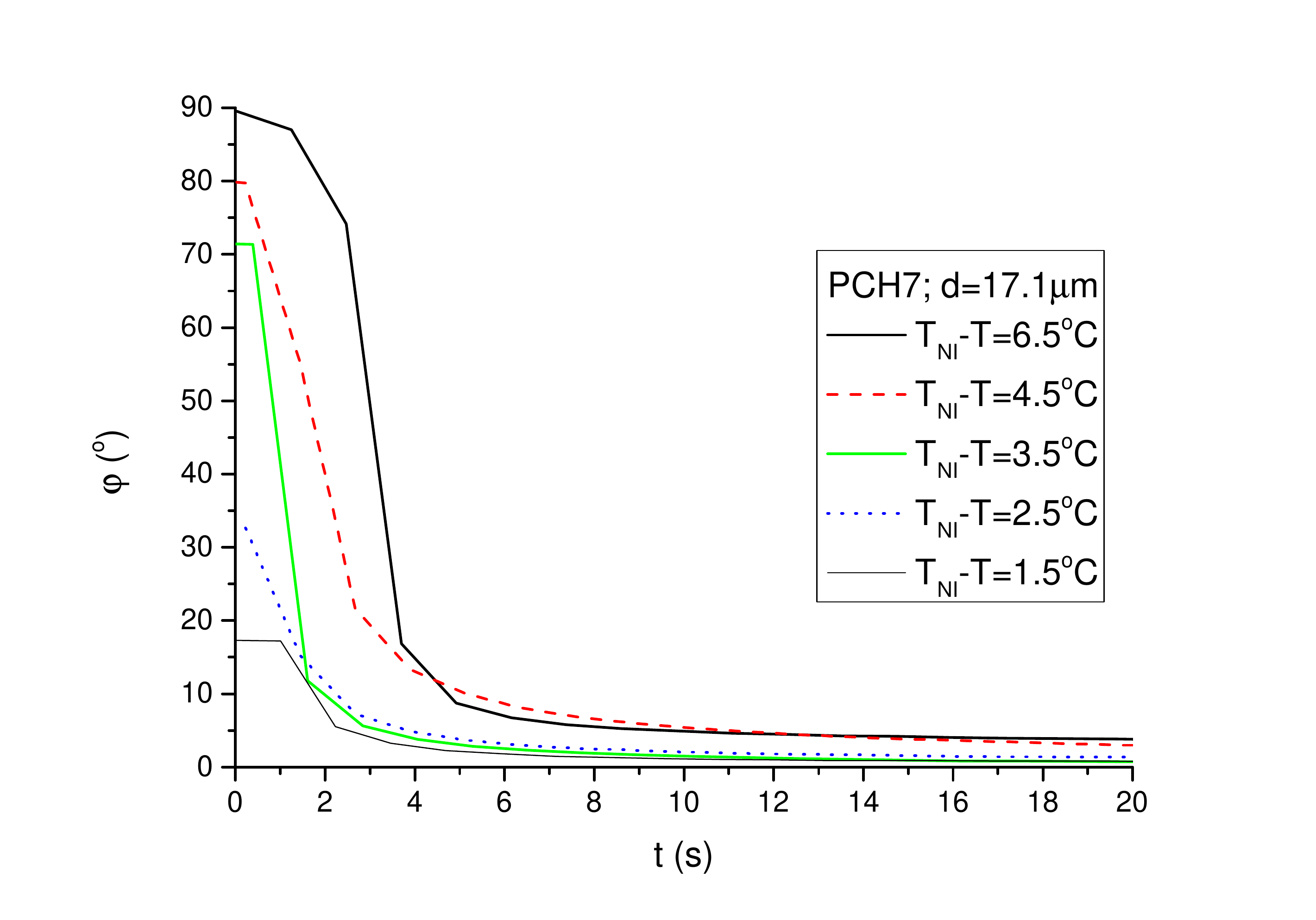}
\caption{Back-relaxation of the azimuthal photo-reorientation angle $\varphi$ (pump beam switched off at $t=0$s) measured at different temperatures as indicated in the legend on a $d=17.1 \mu$m thick cell filled with PCH7. (a) Low-temperature range; (b) mid-temperature range; (c) high-temperature range.} \label{Nassrah_Fig8}
\end{center}
\end{figure}

Regarding the temperature dependence of the relaxation kinetics the statement (and the explanation) made for NLCs with biphenyl rigid core (E7, E63 and 5CB) \cite{Tothkatona2019} also holds for PCH homologue series, as well as for ZLI1695. Namely, with the increase of the temperature the back-relaxation speeds up considerably for all investigated NLCs.
We illustrate the temperature dependence of the back-relaxation kinetics with measurements on PCH7 in Fig. \ref{Nassrah_Fig8}, made in the low-temperature-range (a), mid-temperature-range (b) and in the high-temperature-range (c) of the nematic phase. Obviously, the relaxation times differ by orders of magnitude at high and at low temperatures (from seconds to days).

Finally, from Figs. \ref{Nassrah_Fig3} and \ref{Nassrah_Fig4} one has to note that at a given relative temperature $T_{NI}-T$, the back-relaxation in E7 (and in general, in NLCs having biphenyl in the rigid core -- see in Ref. \cite{Tothkatona2019}) is by orders of magnitude faster than that in ZLI1695, PCH3, PCH5, or in PCH7. To see that, one can compare for example, curve at $T_{NI}-T =32.7^{\circ}$C for E7 [Fig. \ref{Nassrah_Fig3}(a)] with the curve for ZLI1965 at $T_{NI}-T =33.8^{\circ}$C [Fig. \ref{Nassrah_Fig3}(b)], curve at $T_{NI}-T =26.8^{\circ}$C for E7 [Fig. \ref{Nassrah_Fig3}(a)] with the curve for PCH5 at $T_{NI}-T =25.8^{\circ}$C [Fig. \ref{Nassrah_Fig4}(b)], etc.

\section{Discussion}
\label{Discuss}

Measurements presented here (together with the results obtained in Ref. \cite{Tothkatona2019}) have revealed the following new findings on the photoalignment at the pDR1 polymer--NLC interface:

 {\bf (i.)} All investigated NLCs show a close to complete azimuthal photoalignment ($\varphi \approx 90^{\circ}$) at room temperature. The only exception is 5CB, for which $T_{NI}$ is relatively close to room temperature \cite{Tothkatona2019}. As the temperature is increased, at some point $\varphi$ starts to decrease for all NLCs. Depending on the molecular structure of NLCs, considerable differences have been found in the temperature range at which the decrease of $\varphi$ occurs. For NLCs with biphenyl rigid core, the sudden decrease of $\varphi$ has been detected far below $T_{NI}$: for E63 in the temperature range $T_{NI}-T=30-35^{\circ}$C \cite{Tothkatona2019} and for E7 in the range of $T_{NI}-T=20-25^{\circ}$C -- see Fig. \ref{Nassrah_Fig5} and in Ref. \cite{Tothkatona2019}. In contrast to that, in NLCs having phenylcyclohexane or bicyclohexane rigid core the decrease of $\varphi$ has occurred much closer to $T_{NI}$, in the range of $T_{NI}-T \lesssim 10^{\circ}$C -- see Fig. \ref{Nassrah_Fig5}. One has to note that in this high temperature range, a temperature induced change in the pretilt angle (orientational transition towards the homeotropic alignment) has been observed recently for E7 and 5CB NLCs \cite{Tothkatona2019}.

{\bf (ii.)} While in NLCs with biphenyl rigid core a significant zenithal photoalignment has been detected in the temperature range in which the azimuthal photoalignment diminishes \cite{Tothkatona2019}, in NLCs with phenylcyclohexane or bicyclohexane rigid core only indications have been found for zenithal photoalignment of a much smaller magnitude at any temperature of the nematic phase.

{\bf (iii.)} At a given relative temperature $T_{NI}-T$, the back-relaxation in NLCs having biphenyl in the rigid core (E7, E63) is by orders of magnitude faster than that in NLCs with phenylcyclohexane or bicyclohexane rigid core (ZLI1695, PCH3, PCH5, PCH7) -- see Figs. \ref{Nassrah_Fig3}, \ref{Nassrah_Fig4} and in Ref. \cite{Tothkatona2019}.

Findings {\bf (i.)} -- {\bf (iii.)}, namely the diminishing azimuthal photoalignment far below $T_{NI}$ and the appearance of the zenithal photoalignment instead, as well as the fast back-relaxation indicate that in NLCs with biphenyl rigid core something, presumably an interaction between the pDR1 polymer and the NLC, acts against the conventional azimuthal photoalignment and helps the relaxation to the initial (planar) alignment. On the other hand, the persistence of the azimuthal photoalignment to temperatures much closer to $T_{NI}$ (up to temperatures at which presumably the temperature induced orientational transition takes over \cite{Tothkatona2019}), the much smaller (if any) zenithal photoalignment, and the much slower back-relaxation, observed in NLCs with phenylcyclohexane or bicyclohexane rigid core, indicate that in these systems no such interaction is present, or it is much smaller (compared to that in systems with NLCs having biphenyl rigid core).

\begin{figure}
\begin{center}
\includegraphics[width=20pc]{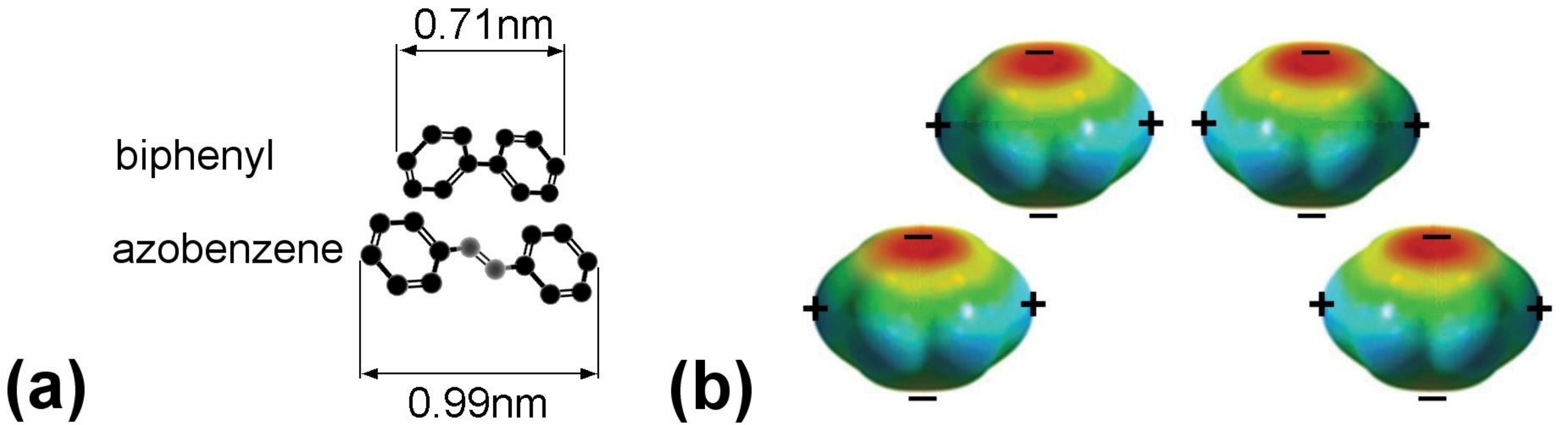}
\caption{(a) Size of the biphenyl rigid core and that of the azobenzene trans-isomer. (b) Offset stacked aromatic interaction between benzene molecules enabled by the molecular electrostatic surface potential (MESP, blue is positive, red is negative).} \label{Nassrah_Fig9}
\end{center}
\end{figure}

To identify the potential source of such interaction, for simplicity, first we consider the size of the biphenyl core \cite{Solak2002,Eelkema2006} and that of the azobenzene trans-isomer \cite{Oscurato2018} only, and we compare them in Fig. \ref{Nassrah_Fig9}(a). Phenyl rings are capable to exhibit aromatic interactions that have been proposed to consist of van der Waals, hydrophobic and electrostatic forces \cite{Hunter2001}. Several geometries have been proposed on the basis of the electrostatic component, arising from interactions of the quadrupole moments of the aromatic rings, based on the greater electron-density on the face of the ring and reduced electron-density on the edge \cite{Hunter1990}. One of them is the so called offset stacked aromatic $\pi - \pi$ interaction \cite{Waters2002,Matthews2014}, which is demonstrated by representation of the molecular electrostatic surface potential (MESP) for four benzene rings in Fig. \ref{Nassrah_Fig9}(b). We note here, that in the biphenyl structure, the two rings are somewhat twisted with respect to each other \cite{Solak2002,Chana2002}, the CN polar head and the alkyl tail in the cyano-biphenyl NLCs \cite{Mandle2015}, as well as the azo-coupling in the azobenzene derivative \cite{Cojocaru2013} introduce minor modifications in the simplified MESP shown in Fig. \ref{Nassrah_Fig9}(b). Despite of these minor modifications, the dimensions of the NLCs with the biphenyl rigid core, and those of the azobenzene-moiety attached to the PMMA polymer are such that considerable offset stacked aromatic $\pi - \pi$ interactions are expected between them. These interactions act against the trans-to-cis isomerization upon the excitation of the azobenzene-moiety (i.e., against the photoalignment) by introducing additional restoring stresses. For further analysis, one has to take into account the temperature dependence of the azimuthal and zenithal anchoring strengths. Based on the experimental results \cite{Tothkatona2019} it has been proposed that the zenithal anchoring strength weakens with the increase of the temperature much faster than the azimuthal anchoring strength. This assumption has been supported by the experimental observations at high temperatures ($T-T_{NI} < 10^{\circ}$C), at which a temperature induced orientational transition from planar towards the homeotropic has been observed without any pump beam illumination \cite{Tothkatona2019}. These phenomena are presumably related to the flexibility of the spacer consisting of two methylene units that connects the PMMA main chain with the azobenzene moiety -- see between the oxygen and nitrogen atoms in Fig. \ref{Nassrah_Fig2}. To our interpretation, under the pump beam illumination, the additional restoring stresses originating from the offset stacked aromatic $\pi - \pi$ interactions in case of the NLCs with biphenyl rigid core may cause the "crossover" between the zenithal and azimuthal anchoring strengths to occur at a considerably lower temperature. In other words, the orientational transition from planar towards homeotropic orientation under pump beam illumination occurs at much lower temperatures than it occurs without the photo-excitation. Naturally, when the photo-excitation is terminated, the driving force for trans-to-cis isomerization is also stopped, and the system relaxes back to the initial planar alignment. Note, that such a scenario does not even necessarily require a traditional trans-to-cis isomerization assisted photoalignment -- the prerequisites are only the stresses induced by the isomerization and by the offset stacked aromatic $\pi - \pi$ interactions, as well as the proper azimuthal and zenithal strengths. To analyze this question in details, future experiments are planned with a variable pump beam light intensity.

On the other hand, one has to note, whenever the pump beam is switched off, the suggested $\pi - \pi$ interaction helps the back-relaxation, which by that becomes faster. This can explain by orders of magnitude faster back-relaxation observed in NLCs with biphenyl rigid core compared to that in NLCs with phenylcyclohexane or bicyclohexane rigid core.

In case of NLCs with phenylcyclohexane or bicyclohexane rigid core such $\pi - \pi$ interaction between the NLC molecules and the azobenzene-moiety of pDR1, acting against the trans-to-cis isomerization is not expected (for the MESP of cyclohexane see e.g. \cite{Emenike2016}). Therefore, the azimuthal photoalignment is not detained; instead it is maintained (almost) over the whole nematic temperature range, the zenithal photoalignment is of a small magnitude, or does not occur at all, and the back-relaxation is much slower than that with NLCs having biphenyl rigid core. Therefore, the suggested $\pi - \pi$ interaction together with the different temperature dependence of the azimuthal and zenithal anchoring strengths (proposed in Ref. \cite{Tothkatona2019}), can give a qualitative explanation for the experimental findings {\bf (i.)} -- {\bf (iii.)} described above.

Besides the findings {\bf (i.)} -- {\bf (iii.)}, there are few more experimental observations worth to discuss.

Focusing on the temperature dependence of the azimuthal photo-reorientation angle $\varphi (\Delta T)$ for PCH homologous series in Fig. \ref{Nassrah_Fig5}, one can note that the azimuthal photoalignment is maintained to temperatures closer to $T_{NI}$ as the length of the alkyl chain is increased. However, it is not straightforward that the length of the alkyl chain directly influences the temperature range of the efficient azimuthal photoalignment. One has to consider other aspects too, for example that the phase transition temperature $T_{NI}$ also increases with the increase of the length of the chain. Therefore, additional measurements on other homologous series are desirable to reveal whether and how the length of the alkyl chain influences photoalignment.

The non-trivial thickness dependence of the back-relaxation kinetics (and that of the temperature range in which the sudden decrease of $\varphi$ is detected for E7 \cite{Tothkatona2019}) indicates that these processes are much more influenced by some condition(s)/factor(s) other than the sample thickness $d$. Such condition/factor could be for example differences in the quality of the pDR1 layer for for different cells (despite of the same preparation procedure), or the exact illumination history with the pump beam for each sample. This open question offers a topic for future experiments.

Finally, we note that we have also performed photoalignment measurements on contaminated, or partially decomposed NLCs of PCH3, PCH5 and Mixture ZLI1132 (a mixture mainly containing PCH homologues: 24\% of PCH3, 36\% of PCH5, 25\% of PCH7, and 15\% of BCH5). The contamination/partial decomposition has been easily detected, since it caused a drastic decrease of the phase transition temperature $T_{NI}$, and a significant broadening of the temperature range where the nematic and isotropic phases coexist: the nematic-to-isotropic phase transition has occurred in the temperature range of $37-41^{\circ}$C, $40-45.5^{\circ}$C, and $59-66^{\circ}$C instead of the expected, literature data $T_{NI}=47^{\circ}$C, $T_{NI}=53.5^{\circ}$C, and $T_{NI}=71^{\circ}$C for PCH3, PCH5 and ZLI1132, respectively. The azimuthal photoalignment angle in all cells prepared with these NLCs has been found negligible ($\varphi < 10^{\circ}$) at any temperature, demonstrating the importance of the quality (purity) of NLCs in contact with the pDR1 polymer for photoalignment.

\section{Conclusions and outlook}
\label{Concl}

We have performed series of experimental studies on the photoalignment at NLC -- photosensitive polymer interface. While employing the same preparation procedure of the photosensitive pDR1 polymer layer, we have systematically varied the rigid core of NLCs from biphenyl, through phenylcyclohexane to bicyclohexane. Substantial differences in the photoalignment process have been found depending on the molecular structure of NLCs. In NLCs with biphenyl rigid core azimuthal photoalignment was detected only in the lower temperature range of the nematic phase. In the higher temperature range of the nematic phase zenithal photoalignment has occured, while at temperatures close to $T_{NI}$ a temperature induced orientational transition (from planar towards the homeotropic) has been also found \cite{Tothkatona2019}. In contrast, in NLCs with phenylcyclohexane or bicyclohexane rigid core no zenithal photoalignment has been detected, and the azimuthal photoalignment has been maintained over almost the whole temperature range of the nematic phase. In these NLCs $\varphi$ decreases only at temperatures close to $T_{NI}$, where temperature induced orientational transition is occuring based on previous measurements \cite{Tothkatona2019}. However, to give a direct, quantitative experimental measure for the temperature induced orientational transition, one needs an NLC for which the temperature dependencies of all relevant material parameters are known as described in Ref. \cite{Tothkatona2019}. Unfortunately, these parameters are unknown for PCH and ZLI1695 NLCs at the present.

We interpret the experimental results with offset stacked aromatic $\pi - \pi$ interactions between the biphenyl rigid core of the NLCs and the azobenzene moiety of the pDR1 polymer.
These interactions are absent in the case of NLCs with bicyclohexane rigid core, or present on a single phenyl ring of NLCs with phenylcyclohexane rigid core and therefore, do not influence the photoalignment process. In the future, the validity of this interpretation can be further tested with NLCs containing biphenyl in the rigid core, but with one of them being modified for example with three fluorene atoms at 1,2,3 (or 3,4,5) positions, which almost "inverts" the MESP of the ring \cite{Shimizu2014}. In that case (similarly to NLCs with phenylcyclohexane), one does not expect zenithal photoalignment, and the azimuthal photoalignment should extend over the (almost) whole nematic temperature range.

Still concentrating on role of the molecular structure of NLCs, the influence of the length of the alkyl chain discussed in the Section \ref{Discuss} can be further clarified in the future for example on the cyanophenyl-dioxane (PDX) homologue series, for which $T_{NI}$ does not increase monotonously with the length of the alkyl chain \cite{Synthon}.

The other future research direction (already mentioned in Ref. \cite{Tothkatona2019}) could be to systematically vary the thickness of the pDR1 layer and/or to change the surface density of the azo-moiety in the pDR1. These investigations could eventually shed light on the non-trivial thickness dependence of the temperature range where the sudden decrease of $\varphi$ occurs, and that of the back-relaxation kinetics discussed in the Section \ref{Discuss}.

\section*{Acknowledgements}

The polymer pDR1 was kindly provided by T. K\'osa and L. Sukhomlinova (Alphamicron Inc., Kent, OH, USA). The authors thank V. Kenderesi for technical help in constructing Fig. \ref{Nassrah_Fig9}. Financial support from the National Research, Development and Innovation Office (NKFIH) Grant No. FK 125134 is gratefully acknowledged.





\end{document}